\documentclass[aps,prc,reprint,floatfix,showpacs,superscriptaddress]{revtex4-1}
\usepackage{amsmath,amssymb,graphicx}
\begin{document}

\hyphenation{Na-ka-tsu-ji Ho-rie Ni-shi-no Ta-gu-chi Ta-ma-sa-ku Ma-e-no Sa-ka-ki-ba-ra pos-i-tron}

\title{\boldmath Quantum Criticality and Inhomogeneous Magnetic Order\\ in Fe-doped $\alpha$-YbAlB$_4$}

\author{D. E. MacLaughlin}
\email[To whom correspondence should be addressed: ]{macl@physics.ucr.edu}
\affiliation{Department of Physics and Astronomy, University of California, Riverside, California 92521, U.S.A.}

\author{K. Kuga}
\altaffiliation{Current address: RIKEN SPring-8 Center, Sayo-gun, Hyogo 679-5148, Japan}
\affiliation{Institute for Solid State Physics, University of Tokyo, Kashiwa 277-8581, Japan}

\author{Lei Shu} 
\affiliation{State Key Laboratory of Surface Physics, Department of Physics,
Fudan University, Shanghai 200433, People’s Republic of China}
\affiliation{Collaborative Innovation Center of Advanced Microstructures,
Fudan University, Shanghai 200433, People’s Republic of China}

\author{O. O. Bernal}
\affiliation{Department of Physics and Astronomy, California State University, Los Angeles, California 90032, U.S.A.}

\author{P. -C. Ho}
\affiliation{Department of Physics, California State University, Fresno, California 93740,U.S.A.}

\author{S. Nakatsuji}
\affiliation{Institute for Solid State Physics, University of Tokyo, Kashiwa 277-8581, Japan}
\affiliation{PRESTO, Japan Science and Technology Agency (JST), 4-1-8 Honcho Kawaguchi, Saitama 332-0012, Japan}

\author{K. Huang}
\author{Z. F. Ding}
\author{C. Tan}
\author{Jian Zhang}
\affiliation{State Key Laboratory of Surface Physics, Department of Physics,
Fudan University, Shanghai 200433, People’s Republic of China}

\date{\today}

\begin{abstract}
The intermediate-valent polymorphs~$\alpha$- and $\beta$-YbAlB$_4$ exhibit quantum criticality and other novel properties not usually associated with intermediate valence. Iron doping induces quantum criticality in $\alpha$-YbAlB$_4$ and magnetic order in both compounds. We report results of muon spin relaxation ($\mu$SR) experiments in the intermediate-valent alloys~$\alpha$-YbAl$_{1-x}$Fe$_x$B$_4$, $x = 0.014$ and 0.25. For $x = 0.014$ we find no evidence for magnetic order down to 25~mK\@. The dynamic muon spin relaxation rate~$\lambda_d$ exhibits a power-law temperature dependence~$\lambda_d \propto T^{-a}$, $a = 0.40(4)$, in the temperature range~100~mK--2~K, in disagreement with predictions by theories of antiferromagnetic (AFM) or valence quantum critical behavior. For $x = 0.25$, where AFM order develops in the temperature range~7.5--10~K, where we find coexistence of meso- or macroscopically segregated paramagnetic and AFM phases, with considerable disorder in the latter down to 2~K.
\end{abstract}


\maketitle

\section{INTRODUCTION}

In certain compounds containing $f$ ions, atomic-like $f$ levels and a wide $s$-$d$ band coexist at the Fermi level. This permits strong admixture of ionic states with differing valence due to hybridization with conduction electrons. Such materials are referred to as \textit{intermediate-valence} (IV), \textit{mixed-valence}, or \textit{valence-fluctuating} compounds. They have a variety of unique thermal and magnetic properties, usually including the ability of even a small admixture of a nonmagnetic valent state to prevent local-moment formation in the ground state~\cite{Varm76}. For nearly-integral valence IV crosses over into Kondo and heavy-fermion behavior~\cite{Hews93}. 

The 4$f$ ions~Ce$^{3+}$ and Yb$^{3+}$, with one $4f$ electron and one hole, respectively, exhibit IV or heavy-fermion behavior (admixture of nonmagnetic Ce$^{4+}$ and Yb$^{2+}$ components, respectively) in many intermetallic compounds. Their properties in Ce- and Yb-based metals are not very symmetric, however; superconductivity, weak-moment magnetism, and quantum criticality are often found in Ce-based compounds but seldom in Yb-based ones. Perhaps more fundamentally, in metals the Ce valence is usually close to 3, whereas Yb ions are more often found in an IV state relatively far from integral valence.

The term \textit{quantum criticality} refers to phenomena involving quantum fluctuations at transitions between phases at $T = 0$. Such effects have been extensively studied in numerous rare-earth-based heavy-fermion metals~\cite{CoSc05,vLRVW07,GSS08}. They include unconventional superconductivity, non-Fermi liquid behavior in the neighborhood of the quantum critical point (QCP), weak-moment antiferromagnetism (AFM), quasi-ordered phases such as `spin nematics,' and even more exotic phases involving modification of the fundamental nature of the electrons involved. Such phenomena are associated with the interplay between magnetic interactions and local-moment screening by the Kondo effect and its heavy-fermion cousin, both of which are found near integral valence; quantum criticality has seldom been searched for in IV materials.

The polymorphs $\alpha$-YbAlB$_4$ and $\beta$-YbAlB$_4$~\cite{MNKT07,NKMT08,KKMM08,MKTK11} and their alloys with iron~\cite{KMTC12,KuNa13,SIKK13,KMSO16u} display a rich variety of unexpected properties, and promise to shed light on a number of interesting phenomena. They are both substantially intermediate-valent (Yb$^{z+}$, $z = 2.73$ and 2.75 for $\alpha$-YbAlB$_4$ and $\beta$-YbAlB$_4$, respectively)~\cite{OMIE10} but, very surprisingly, retain local-moment behavior to low temperatures~\cite{NKTM10,MKTK11}. $\beta$-YbAlB$_4$ is one of the few pure rare-earth-based materials to exhibit quantum criticality without tuning, i.e., without doping, pressure, or magnetic field, as evidenced by the scaling of the temperature and magnetic field dependence of magnetization~\cite{MNKK11}. Applied magnetic fields rapidly restore Fermi-liquid behavior. $\beta$-YbAlB$_4$ is also the only known Yb-based heavy-fermion superconductor: high-purity crystals are superconducting below $T_c \approx 0.08$~K~\cite{NKMT08,KKMM08}. The superconductivity evolves from the quantum-critical state and is very fragile, appearing only for samples with low residual resistivities. This strong sensitivity of $T_c$ to sample purity suggests that the superconductivity is of an unconventional, non-$s$-wave type~\cite{KKMM08}. 

Undoped $\alpha$-YbAlB$_4$ is not a quantum critical system (no divergence of $C_p/T$), but the solid solution~$\alpha$-YbAl$_{1-x}$Fe$_x$B$_4$ can be tuned to quantum criticality at a critical concentration~$x_\mathrm{cr} = 0.014$~\cite{KMTC12,KMSO16u}. There is evidence from thermodynamic and photoemission data that valence fluctuations are involved in the quantum critical behavior~\cite{KMSO16u}. More heavily Fe-doped samples exhibit a first-order transition to a canted antiferromagnetic (AFM) phase. In particular, magnetization and M\"ossbauer-effect measurements on $\alpha$-YbAl$_{0.75}$Fe$_{0.25}$B$_4$~\cite{SIKK13,SIKN14a} show evidence for a complex phase transition; the magnetization exhibits anomalies at 9.4~K, 8.0~K and 6.9~K~\cite{SIKN14a} that have been attributed to magnetic ordering. $^{57}$Fe M\"ossbauer experiments~\cite{SIKK13} and the absence of magnetism in Fe-doped LuAlB$_4$~\cite{KMTC12} confirm that the doped Fe is itself nonmagnetic; the static magnetism is due to Yb moments only.

We have used the muon spin relaxation ($\mu$SR) technique~\cite{Sche85,Brew94,YaDdR11} to study polycrystalline samples of $\alpha$-YbAl$_{1-x}$Fe$_x$B$_4$, $x = 0.014$ and 0.25. Our goals were to examine the muon spin dynamic (spin-lattice) relaxation in the $x = 0.014$ sample for evidence of the putative quantum critical point, and to search for magnetic transitions in both samples via the onset of a static field or distribution of static fields. Experiments were carried out in zero applied field (ZF) over the temperature range~0.025--15~K, and in weak longitudinal fields (LF) (i.e., field parallel to the initial muon spin direction) at selected temperatures in this range. 

For $x = 0.014$ no evidence was found for static magnetism $\gtrsim 10^{-2}\mu_B$/Yb ion down to 25~mK\@. In this sample the dynamic muon spin relaxation rate~$\lambda_d$ is found to obey a power-law temperature dependence: $\lambda_d(T) \propto T^{-a}$ above 100 mK, with $a = 0.40(4)$ and a maximum in the neighborhood of 50~mK\@. This indicates a divergent density of magnetic excitations (with a possible cutoff near the zero of energy), apparently associated with the QCP at $x = x_\mathrm{cr}$. Such a divergence does not agree with theoretical results based on either AFM or valence quantum criticality~\cite{[{For a review, see }] MiWa14}, both of which yield negative values of $a$. The divergence is consistent with a ferromagnetic (FM) instability~\cite{IsMo96}, which, however, would not account for the results of other experiments noted above. More work is necessary to resolve this discrepancy. 

In $\alpha$-YbAl$_{0.75}$Fe$_{0.25}$B$_4$ the onset of static magnetism over a transition region from 7.5 to 10~K is clearly seen in ZF-$\mu$SR relaxation as a wide distribution of local magnetic fields. The data are consistent with an inhomogeneous distribution of two phases, AFM and paramagnetic (PM), in the transition region, and there are indications of multiple transitions. The fraction of PM phase decreases to zero below $\sim$8~K\@. In the AFM phase the local field is widely distributed, with no signature of a well-defined nonzero average. In this sample $\lambda_d$ exhibits a broad maximum at $\sim$8.5~K suggestive of dynamic critical slowing down of Yb moment fluctuations, and becomes constant below $\sim$6~K\@.

\section{EXPERIMENT}

Flux-grown small crystals of $\alpha$-YbAl$_{1-x}$Fe$_x$B$_4$, $x = 0.014$ and 0.25. were prepared as described previously~\cite{MNKT07}. They were characterized using powder x-ray diffraction and magnetization measurements. 

$\mu$SR experiments were carried out at TRIUMF, Vancouver, Canada, using the $\mu$SR dilution refrigerator at the M15 muon beam line for the temperature range 25~mK--2.5~K\@. The LAMPF $\mu$SR spectrometer at the M20C beam line was used for temperatures between 2~K and 300~K\@.  Data were analyzed using the Paul Scherrer Institute \textsc{musrfit} fitting program~\cite{SuWo12} and the TRIUMF \textsc{physica} programming environment~\footnote{\mbox{\tt http://computing.triumf.ca/legacy/physica/}}.

For time-differential $\mu$SR in solids positive muons ($\mu^+$) are normally used~\footnote{Positive muons occupy interstitial crystalline sites, and thus reflect electronic magnetism better than negative muons that are bound to nuclei in small hydrogenic orbitals~\protect~\cite{Sche85,Brew94,YaDdR11}.}. The time evolution of the decay positron count rate asymmetry~$A(t)$ is proportional to the total (sample plus background) $\mu^+$ spin polarization~$P_\mathrm{tot}(t)$:
\begin{equation} \label{eq:asy}
A(t) = A_0P_\mathrm{tot}(t) \,,
\end{equation}
where the initial asymmetry~$A_0$ is spectrometer-depen\-dent but is usually $\sim 0.2$. The observed asymmetry often contains a component due to muons that miss the sample and stop elsewhere in the spectrometer. In the following this signal is subtracted, and the data are normalized by $A_0$ to yield the ensemble spin polarization~$P(t)$ in the sample.

Two categories of processes contribute to the relaxation of $P(t)$: \textit{static} relaxation, due to an inhomogeneous distribution of time-average local fields~$\langle B_\mathrm{loc}\rangle$ at $\mu^+$ sites, and \textit{dynamic} relaxation, due to thermal fluctuations~$\delta B_\mathrm{loc}(t)$ of the $\mu^+$ local fields around their time averages. Static relaxation is due to (quasistatic) nuclear dipolar fields in dia- and paramagnets, and to coupling to static magnetism if present. 

Dynamic relaxation usually arises from coupling to electronic spin fluctuations~\footnote{For convenience we use the term ``spin fluctuations'' to refer to any fluctuating electronic magnetism, whether or not orbital magnetism is involved.}. If the fluctuation rate~$1/\tau_c$ is in the so-called motional narrowing limit~$\gamma_\mu \langle \delta B_\mathrm{loc}^2\rangle^{1/2}\tau_c \ll 1$~\cite{Abra61,Slic96}, the resulting $\mu^+$ spin polarization can be modeled by
\begin{equation} \label{eq:expdamp}
P(t) = e^{-\lambda_dt}G_s(t) \,, \quad \lambda_d \approx \gamma_\mu^2 \langle \delta B_\mathrm{loc}^2\rangle\tau_c \,,
\end{equation}
where $G_s(t)$ is the appropriate static relaxation function. We expect situations of this kind in the present study, and are thus motivated to fit forms of Eq.~(\ref{eq:expdamp}) to the data.

\section{RESULTS AND DISCUSSION} \label{sec:results}

\subsection{\boldmath $\alpha$-YbAl$_{1-x}$Fe$_x$B$_4$, $x = 0.014$} \label{sec:Fe1.4}

\subsubsection{Zero-field $\mu$SR} \label{sec:Fe1.4ZF}

\begin{figure}[ht]
\includegraphics[clip=,width=3.25in]{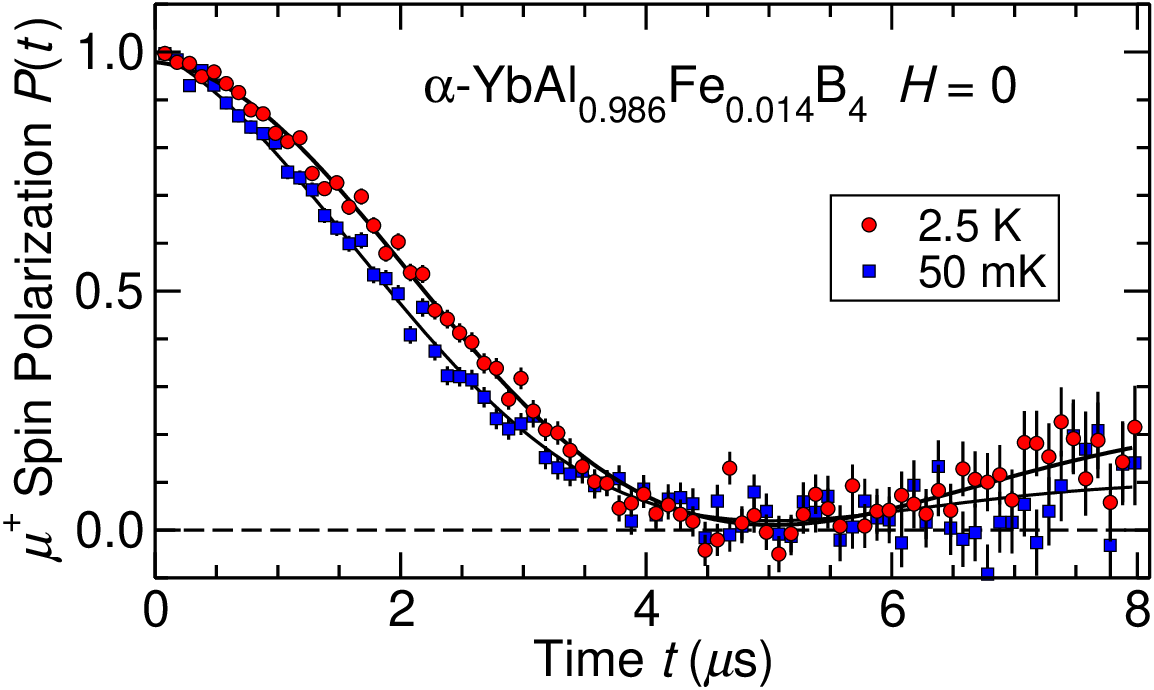}%
\caption{\label{fig:K-Fe11-ZF-pol}(Color online) Zero-field (ZF) $\mu^+$ spin relaxation in $\alpha$-YbAl$_{0.986}$Fe$_{0.014}$B$_4$, $T = 2.5$~mK (circles) and 50~mK (squares). Curves: fits of exponentially-damped ZF Gaussian Kubo-Toyabe function [Eq.~(\protect\ref{eq:edzfgkt})] to the data.}
\end{figure}

Figure~\ref{fig:K-Fe11-ZF-pol} shows $P(t)$ for $\alpha$-YbAl$_{0.986}$Fe$_{0.014}$B$_4$ at 2.5~K and 50~mK in zero field (ZF).
The curves are fits to the data of the exponentially-damped relaxation function
\begin{equation} \label{eq:edzfgkt}
P(t) = e^{-\lambda_dt}G_\text{G}(t) \,,
\end{equation}
where 
\begin{equation} \label{eq:zfgkt}
G_\text{G}(t) = \frac{1}{3} + \frac{2}{3}(1 - \Delta^2t^2)\exp\left(-{\textstyle\frac{1}{2}}\Delta^2t^2\right)
\end{equation}
is the ZF Gaussian Kubo-Toyabe (KT) function~\cite{KuTo67,*HUIN79} appropriate to relaxation by a randomly-oriented Gaussian distribution of static local fields. The relaxation rate~$\Delta$ is the rms width $\Delta/\gamma_\mu$ of the local field distribution in ``frequency units''. The data exhibit the minimum in $P(t)$ and recovery at late times associated with Eq.~(\ref{eq:zfgkt})~\cite{KuTo67,*HUIN79}. It can be seen that there is a small but measurable increase in relaxation rate at low temperature, together with a change in shape of $P(t)$ associated with an increase of $\lambda_d$ relative to $\Delta$.

In the filled skutterudite compound~PrPt$_4$Ge$_{12}$, combined Gaussian and exponential relaxation has been reported~\cite{MSKG10} for which a Lorentzian component of the static field distribution rather than dynamic spin fluctuations is mainly responsible. We therefore consider a generalization of the ZF Gaussian KT relaxation function to the case of a combined Gaussian and Lorentzian static field distribution, the so-called ZF Voigtian static KT function~\cite{MSKG10}:
\begin{equation} \label{eq:zfvkt}
G_V(t) = \frac{1}{3} + \frac{2}{3}(1 - \lambda t - \Delta^2t^2)\exp\left(-\lambda t - {\textstyle\frac{1}{2}}\Delta^2t^2\right) \,.
\end{equation}
The shape of the relaxation function is controlled by the ratio~$\Delta/\lambda$: the limit~$\lambda \to 0$ yields Eq.~(\ref{eq:zfgkt}), whereas the limit $\Delta \to 0$ yields the ZF exponential KT function appropriate to dilute local-moment systems with $1/r^3$ interactions with the muon~\cite{UYHS85}. Equation~(\ref{eq:zfvkt}) should be considered an empirical interpolation between the Gaussian and exponential limits.

Fits of the exponentially-damped ZF Voigtian KT function 
\begin{equation} \label{eq:edzfvkt}
P(t) = e^{-\lambda_d t}G_V(t)
\end{equation}
to the data for $x = 0.014$ (not shown) yield $\lambda \approx 0$ (and $\lambda_d \ne 0$); there is no evidence for static exponential relaxation in this sample. We shall see in Sec.~\ref{sec:Fe25}, however, that in the high-temperature PM phase of $\alpha$-YbAl$_{0.75}$Fe$_{0.25}$B$_4$ fits to ZF data using Eq.~(\ref{eq:edzfvkt}) yield nonzero $\lambda$ (and $\lambda_d \approx 0$).

\begin{figure}[ht]
\includegraphics[clip=,width=3.25in]{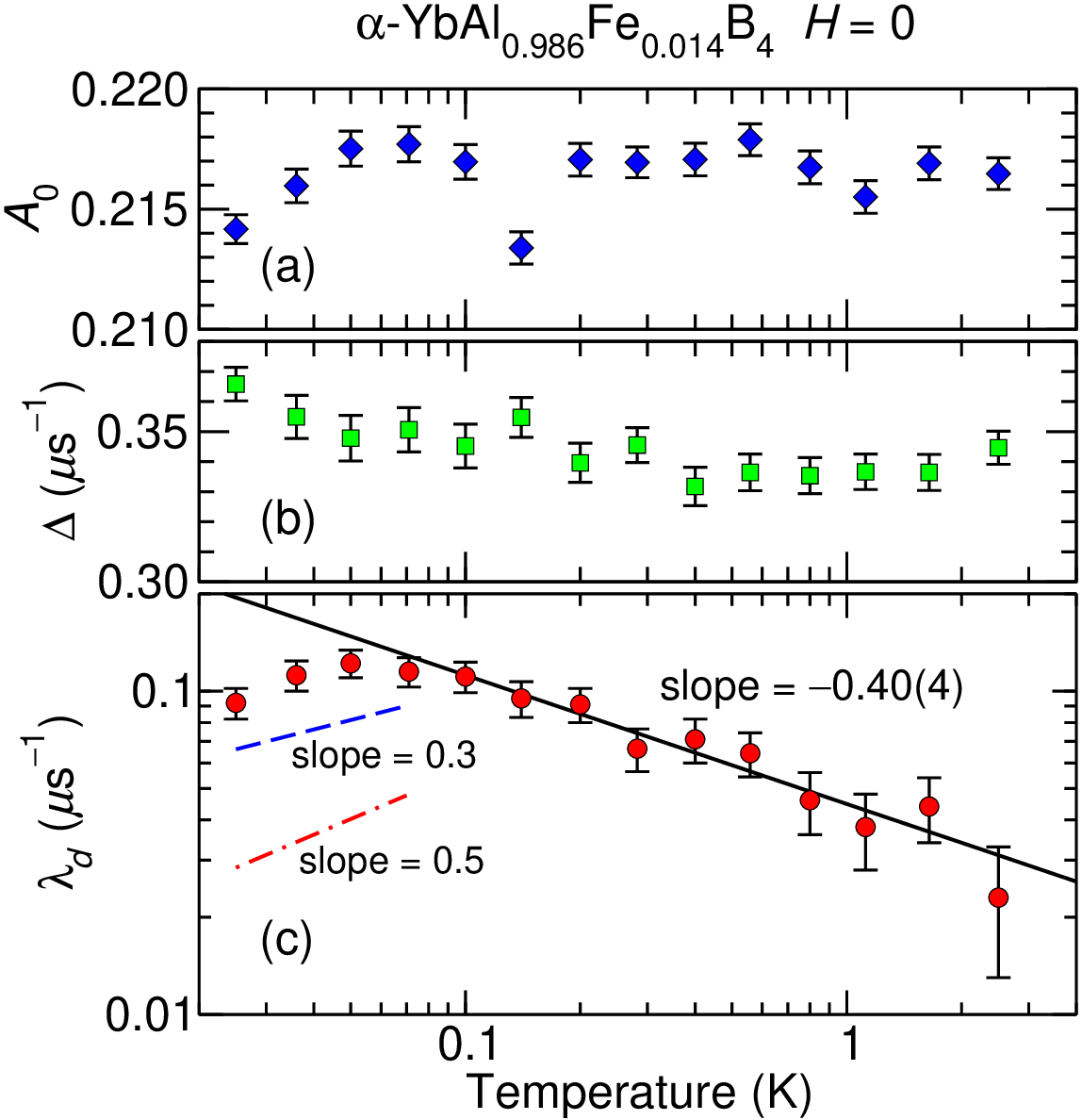}%
\caption{\label{fig:K-Fe18_ZF-rates}(Color online) Temperature dependences of ZF $\mu^+$ initial count rate asymmetry and spin relaxation rates in $\alpha$-YbAl$_{0.986}$Fe$_{0.014}$B$_4$. (a)~Initial asymmetry~$A_0$. (b)~Static Gaussian KT rate~$\Delta$. (c)~Dynamic relaxation rate~$\lambda_d$. Solid line: fit of the power law~$\lambda_d \propto T^{-\alpha}$ to the data for $T \ge 100$~mK\@. The dashed and dash-dot lines represent the range of slopes predicted by the theory of Ref.~\cite{MiWa14}.}
\end{figure}

Figure~\ref{fig:K-Fe18_ZF-rates} gives the ZF temperature dependences of the $t{=}0$ asymmetry~$A_0$ [Eq.~(\ref{eq:asy})], the static KT relaxation rate~$\Delta$, and the dynamic rate~$\lambda_d$ for temperature~$T$ between 25~mK and 2.5~K\@.
Over this range $A_0$ and $\Delta$ are constant to within a few percent of their averages [Figs.~\ref{fig:K-Fe18_ZF-rates}(a) and \ref{fig:K-Fe18_ZF-rates}(b), respectively]. There is no sign of oscillations that would indicate a well-defined static field, and there is no ``missing asymmetry'' from very rapid relaxation due to a strong magnetic transition. The near constancy of $\Delta$ [average value~$0.345(9)~\mu\mathrm{s}^{-1}$] indicates no static magnetism at the level of ${\sim}0.01\mu_B$ per unit cell. These results rule out the onset of static magnetism above 25~mK\@. The value of $\Delta/\gamma_\mu$ is roughly consistent with $^{171}$Yb, $^{27}$Al, and $^{11}$B nuclear dipolar fields. A quantitative comparison would require knowledge of the $\mu^+$ stopping site, which is not known at present.

In contrast, there is a significant increase of $\lambda_d$ with decreasing temperature, followed by a broad maximum at $\sim$50~mK\@. Above 100~mK the data follow a power law~$\lambda_d \propto T^{-a}$, with $a = 0.40(4)$. This divergence followed by a maximum suggests the onset of quantum ($T = 0$) critical spin fluctuations with a cutoff at low frequencies. 

A divergent $\lambda_d(T)$ is, however, not predicted by theories of either AFM or valence criticality~\cite{MiWa14}. The latter has been proposed as a mechanism for quantum critical phenomena in a number of Ce- and Yb-based heavy-fermion compounds including $\alpha$- and $\beta$-YbAlB$_4$~\cite{KMSO16u}. The dynamic relaxation rate ($1/T_1$ in NMR terminology) of a spin probe (nuclear or muon spin) has been calculated within this theory, and vanishing of $1/T_1(T)$ as $T \to 0$ is obtained: $1/T_1 = \lambda_d \propto T^{0.3\text{--}0.5}$, in marked disagreement with the data above 100~mK (Fig.~\ref{fig:K-Fe18_ZF-rates}). AFM spin fluctuations also result in $a < 0$, i.e., vanishing $1/T_1$ as $T \to 0$~\cite{IsMo96,MYI09,MiWa14}.

It is possible that the comparison should be made at lower temperatures, below the maximum in Fig.~\ref{fig:K-Fe18_ZF-rates}(c). The available temperature range down to the cryostat base temperature of 25~mK is too limited for a quantitative comparison, but the data are consistent with the theoretically expected~\cite{MiWa14} range of slopes (dashed and dash-dot lines in Fig.~\ref{fig:K-Fe18_ZF-rates}). This would restore agreement with the valence criticality scenario. It would, however, leave the origin of the power law above 100~mK unexplained.

Power-law temperature dependences of the $\mu^+$ dynamic relaxation rate have been observed in a number of systems that exhibit the non-Fermi liquid behavior often associated with quantum criticality. These include CeP$_{0.15}$Rh$_{0.85}$~\cite{AHPK08}, YbCu$_{5-x}$Au$_x$, $x = 0.6$~\cite{CPGB09}, and YbNi$_4$P$_2$~\cite{SGKY12}. In these cases the exponent~$a$ varies between 0.3 and 0.8. The divergence has been taken as a sign of a FM QCP, primarily on the basis of the qualitative agreement with predictions of the self-consistent renormalization (SCR) theory~\cite{IsMo96} for FM criticality. For AFM criticality SCR theory predicts a negative value of $a$, as does a later proposal of quantum tricriticality~\cite{MYI09}. Magnetization measurements~\cite{KMTC12} exhibit hysteresis along the $c$ axis, and suggest a FM component of the ordered magnetization in the $ab$ plane of $\alpha$-YbAl$_{1-x}$Fe$_x$B$_4$, $x > x_\mathrm{cr}$. Fluctuations associated with this component could dominate the $\mu^+$ dynamic relaxation for $x = x_\mathrm{cr}$. However, Ref.~\onlinecite{IsMo96} predicts a maximum in $1/T_1$ at low temperatures, associated with coupling between spin fluctuation modes around the critical wave vector. This is consistent with the data [Fig.~\ref{fig:K-Fe18_ZF-rates}(c)], but parameter values necessary for quantitative comparison are not known. In any case, bulk properties of $\alpha$-YbAl$_{0.986}$Fe$_{0.014}$B$_4$~\cite{KMSO16u} are not consistent with the FM QCP scenario~\cite{[{See, e.g., }]Stew01}.

\subsubsection{Longitudinal-field $\mu$SR} \label{sec:Fe1.4LF}

\begin{figure}[ht]
\includegraphics[clip=,width=3.25in]{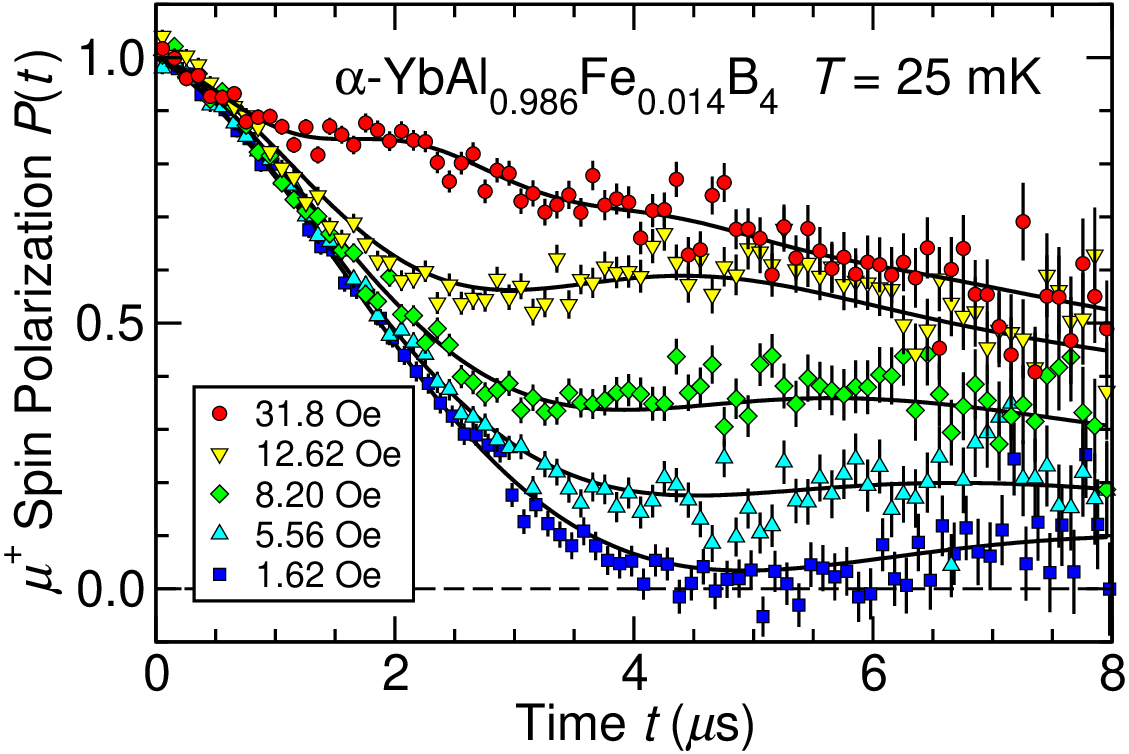}%
\caption{\label{fig:K-Fe11-25mK-pol}(Color online) Longitudinal-field (LF) $\mu^+$ spin relaxation in $\alpha$-YbAl$_{0.986}$Fe$_{0.014}$B$_4$, $T = 25$~mK\@. Curves: fits to the data of exponentially-damped static LF Gaussian Kubo-Toyabe function~\protect~\cite{HUIN79}.}
\end{figure}

The dependence of $P(t)$ on longitudinal field (LF) in $\alpha$-YbAl$_{0.986}$Fe$_{0.014}$B$_4$ at 25~mK is shown in Fig.~\ref{fig:K-Fe11-25mK-pol}.
As in ZF, the data are well fit by an exponentially-damped static relaxation function [Eq.~(\ref{eq:expdamp})], where in this case $G_s(t)$ is the static Gaussian KT relaxation function in nonzero LF~\cite{HUIN79}. The majority of the field dependence is due to ``decoupling'' of the muon spin from random static internal fields by the longitudinal field~$H_L$ for $ H_L \gtrsim \Delta/\gamma_\mu$. For $H_L = 31.8$~Oe (Fig.~\ref{fig:K-Fe11-25mK-pol}) the decoupling is nearly complete and the relaxation is mainly dynamic~\cite{HUIN79}. 

\begin{figure}[ht]
\includegraphics[clip=,width=3.25in]{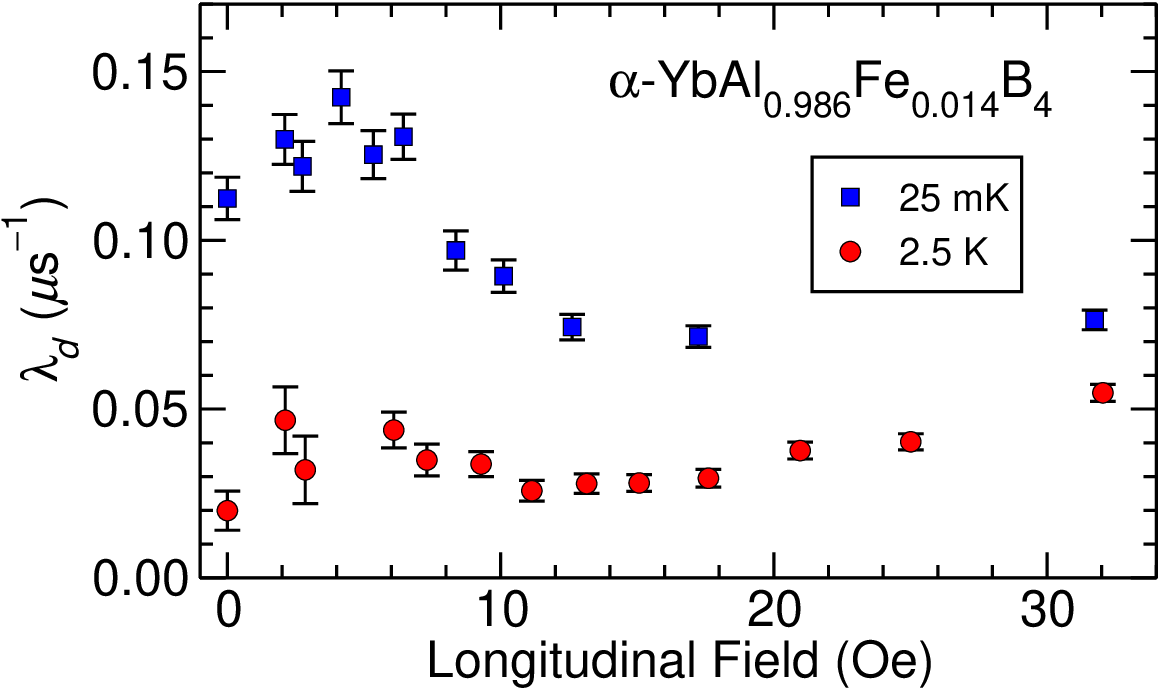}%
\caption{\label{fig:K-Fe18_LF-rates}(Color online) Dependence of dynamic $\mu^+$ spin relaxation rate~$\lambda_d$ on longitudinal field in $\alpha$-YbAl$_{0.986}$Fe$_{0.014}$B$_4$ at 25~mK (squares) and 2.5~K (circles).}
\end{figure}

The relaxation rate~$\lambda_d$ varies considerably with field, as shown in Fig.~\ref{fig:K-Fe18_LF-rates} for $T = 25$~mK and 2.5~K\@.
At both temperatures $\lambda_d$ goes through a maximum at $\sim$4~Oe, followed by a shallow minimum at somewhat higher fields. It is hard to see how such weak fields could modify the electronic spin system significantly. 

This field dependence is reminiscent of that observed in Cu metal, which was attributed to avoided level crossing (ALC) of muon Zeeman and nuclear quadrupolar energy levels~\cite{KBHK86}. In ALC the maximum in $\lambda_d$ occurs at roughly $\omega_Q/\gamma_\mu$, where $\omega_Q$ is the nuclear quadrupolar splitting frequency. However, $\omega_Q/2\pi$ obtained from the peak field is $\sim$0.05~MHz, which is an order of magnitude smaller than values obtained from quadrupole-split $^{11}$B NMR in $\alpha$- and $\beta$-YbAlB$_4$~\cite{TGKY16}. It seems unlikely either that (1)~the muon spin couples predominantly to $^{27}$Al or $^{173}$Yb (NMR has not yet been reported for either of these nuclei, and $^{173}$Yb is only 16\% abundant), or (2)~the small concentration of iron dopant or the additional contribution of the muon electric field gradient~\cite{KBHK86} cancels the intrinsic crystalline contribution to $\omega_Q(^{11}\mathrm{B})$ to this degree. Thus the origin of the observed field dependence remains uncertain.

\subsection{\boldmath $\alpha$-YbAl$_{1-x}$Fe$_x$B$_4$, $x = 0.25$} \label{sec:Fe25}

(For convenience we refer to magnetic order in this system as ``antiferromagnetic'' or ``AFM'', in spite of the evidence for FM criticality discussed in Sec.~\ref{sec:Fe1.4ZF}.)

The behavior of the $\mu^+$ relaxation for $x = 0.25$ can be divided into three temperature regions: (1)~a fully PM region~$T \gtrsim 10$~K, (2)~a fully AFM region~$T \lesssim 7.5$~K, and (3)~a transition region between these temperatures. In all three regions the damped ZF Gaussian KT function [Eq.(\ref{eq:edzfgkt})] give poor fits to ZF data, whereas for the PM and AFM regions damped ZF Voigtian KT fits [Eq.(\ref{eq:edzfvkt})] are statistically satisfactory. This is evidence for a local field distribution function with a Lorentzian component, i.e., with more weight in the ``wings'' or ``shoulders'' than for a purely Gaussian distribution.

In the transition region neither Gaussian nor Voigtian KT functions give satisfactory fits, but good fits were obtained to a sum of PM and AFM relaxation functions, with an AFM fraction~$f_\mathrm{AFM}$ that decreases monotonically from $f_\mathrm{AFM} = 1$ at $\sim$8~K to 0 at $\sim$10~K\@. This indicates that the transition region is macroscopically inhomogeneous.

We first consider data from the PM and AFM temperature regions.

\subsubsection{Voigtian and power-exponential relaxation functions} \label{sec:sv}

 An alternative to the Voigtian KT relaxation function for interpolation between Gaussian and exponential KT relaxation functions is provided by the ZF power exponential (PE)~\cite{CrCy97}
\begin{equation} \label{eq:edzfskt}
G_\mathrm{PE}(t) = \frac{1}{3} + \frac{2}{3}[1 - (\sigma t)^\beta] \exp[-(\sigma t)^\beta/\beta] \,.
\end{equation}
The exponential and Gaussian KT relaxation functions are limits for $\beta = 1$ and 2, respectively. The shape of the PE relaxation function is controlled by $\beta$, in a manner analogous to the ratio~$\Delta/\lambda$ for the Voigtian (Sec.~\ref{sec:Fe1.4ZF}). 

\begin{figure}[ht]
\includegraphics[clip=,width=3.25in]{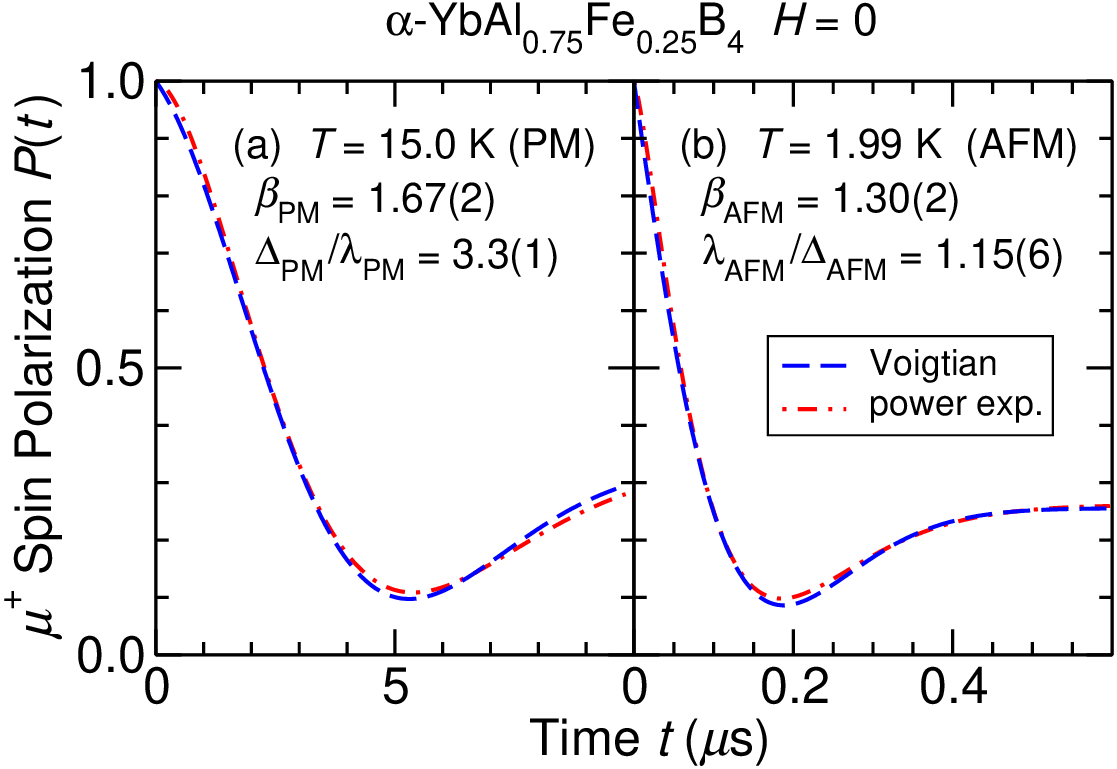}%
\caption{\label{fig:sv-comp}(Color online) Voigtian [Eq.(\ref{eq:edzfvkt}), dashed curves] and power-exponential [Eq.(\ref{eq:edzfskt}), dash-dot curves] Kubo-Toyabe relaxation functions from fits to ZF $\mu^+$ spin polarization data from $\alpha$-YbAl$_{0.75}$Fe$_{0.25}$B$_4$. (a)~PM phase, $T = 15.0$~K\@. (b)~AFM phase, $T = 1.99$~K\@.}
\end{figure}

Both Voigtian and PE functions have been used when a more exact model of the field distribution is unavailable or cumbersome~\cite{YaDdR11}. Thus it is useful to examine whether or not for some intermediate distributions the data would be better fit by one or the other interpolating function. This is done by comparing Voigtian and PE fits to data from YbAl$_{0.75}$Fe$_{0.25}$B$_4$ at temperatures well above and well below the AFM transition. Figure~\ref{fig:sv-comp} shows the comparison~\footnote{In Fig.~\ref{fig:sv-comp}(b) the fit curves include the effect of preferential crystallite orientation discussed in Sec.~\ref{sec:AFM}.}. For clarity only the fits are shown; the data are discussed below. 
By eye the curves of Fig.~\ref{fig:sv-comp} seem more nearly Gaussian in the PM phase and more nearly exponential in the AFM phase. The fit values of the parameters~$\beta$ (PE fits) and $\Delta/\lambda$ (Voigtian fits) confirm this qualitative impression: for the PE fits $\beta_\mathrm{PM}$ is significantly larger than $\beta_\mathrm{AFM}$, and for the Voigtian fits $\Delta_\mathrm{PM} > \lambda_\mathrm{PM}$ and $\Delta_\mathrm{AFM} < \lambda_\mathrm{AFM}$ in the PM and AFM states, respectively. 

It can be seen that the Voigtian and PE functions are very similar, and there is no significant difference between them in goodness of fit. There is, however, one situation in which the PE fit is more flexible, viz., if there is even more weight in the shoulders than for a Lorentzian field distribution. A PE fit can accommodate this with a value of $\beta$ less than 1 (a ``stretched exponential''), whereas a Voigtian fit only interpolates between the exponential and Gaussian limits. We shall see in Sec.~\ref{sec:trans} that exponentially-damped PE fits in the transition region near 10~K yield $\beta < 1$, and we therefore use this function for fits in the AFM phase. For fits in the PM phase we have arbitrarily chosen the exponentially-damped Voigtian KT function [Eq.~(\ref{eq:edzfvkt})].

\subsubsection{Paramagnetic Phase} \label{sec:par}

\begin{figure}[ht]
\includegraphics[clip=,width=3.25in]{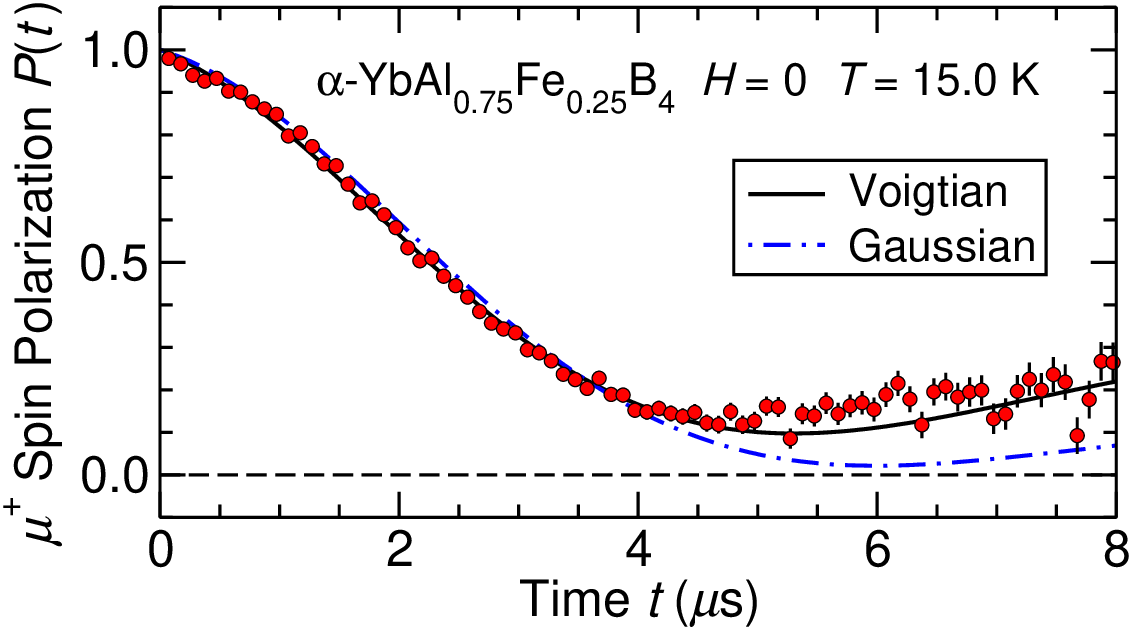}%
\caption{\label{fig:par-Voigt}(Color online) $\alpha$-YbAl$_{0.75}$Fe$_{0.25}$B$_4$ $\mu$SR ZF asymmetry time spectrum, $T = 15.0$~K\@. Solid curve: exponentially-damped Voigtian KT relaxation [Eq.~(\protect\ref{eq:edzfvkt}), $\lambda_d \approx 0$]. Dash-dot curve: exponentially-damped Gaussian KT relaxation [Eq.~(\protect\ref{eq:edzfgkt})].}
\end{figure}

Figure~\ref{fig:par-Voigt} shows the time evolution of the $\mu^+$ spin polarization~$P(t)$ in $\alpha$-YbAl$_{0.75}$Fe$_{0.25}$B$_4$, $T = 15.0$~K\@.
The $\mu^+$ data are similar to those for the $x = 0.014$ sample at high temperatures (cf.\ Fig.~\ref{fig:K-Fe11-ZF-pol}). The solid curve is a fit to the exponentially-damped Voigtian KT relaxation function given by Eq.~(\ref{eq:edzfvkt}). This fit yields PM-phase static relaxation rates~$\Delta_\mathrm{PM} = 0.291(3)~\mu\mathrm{s}^{-1}$ and $\lambda_\mathrm{PM} = 0.087(4)~\mu\mathrm{s}^{-1}$ [Eq.~(\ref{eq:zfvkt})], and dynamic rate~$\lambda_d \ll 0.01~\mu\mathrm{s}^{-1}$ [Eq.~(\ref{eq:edzfvkt})]. For comparison, the dashed curve (which is not a fit) gives the exponentially damped Gaussian KT function of Eq.~(\ref{eq:edzfgkt}) with the same value of $\Delta_\mathrm{PM}$ and $\lambda_d = 0.087~\mu\mathrm{s}^{-1}$. 

The latter curve agrees with the former and with the data only at early times ($\lesssim 3~\mu\mathrm{s}$). At late times the Voigtian function without damping return to the value~$1/3$ as generally expected~\cite{Sche85,YaDdR11} for static relaxation only and randomly-oriented local fields. This return is in better agreement with the data than the overall damping imposed by Eq.~(\ref{eq:edzfgkt}). 

\begin{figure}[ht]
\includegraphics[clip=,width=3.25in]{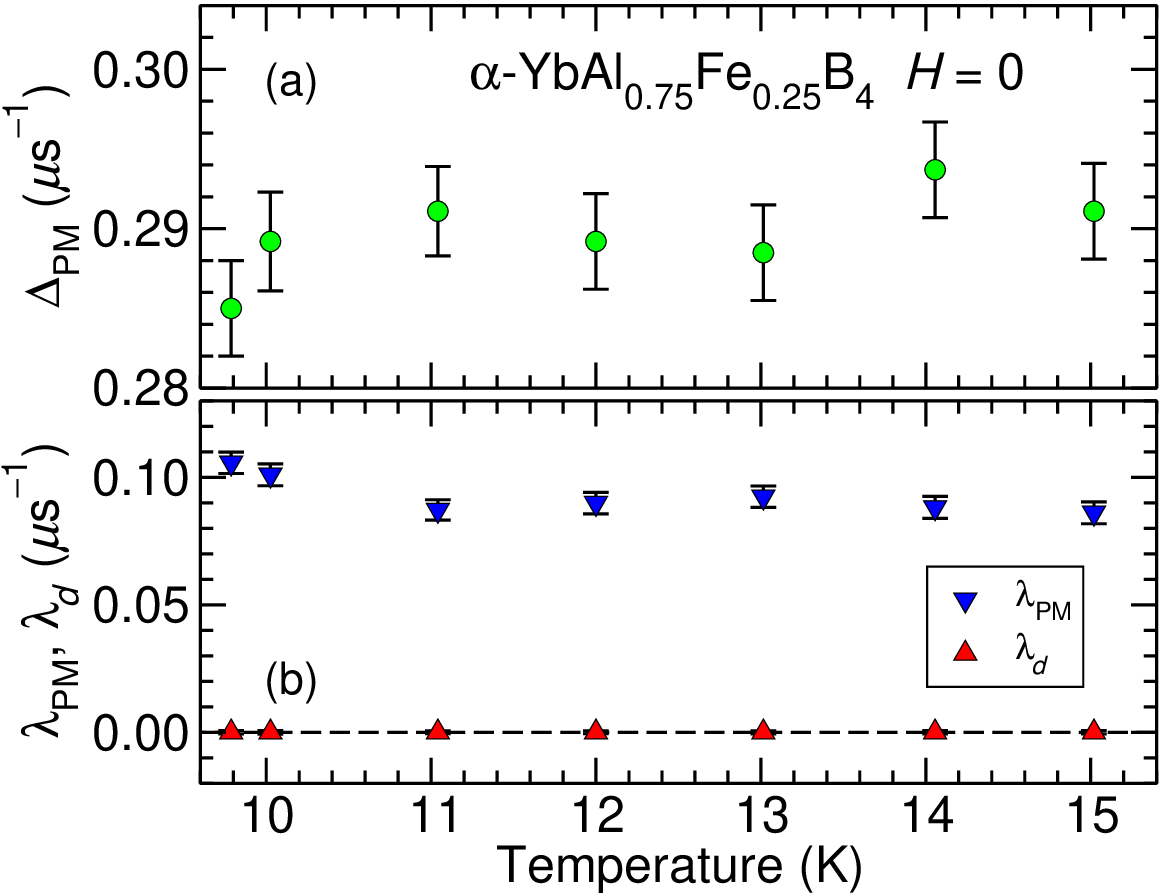}%
\caption{\label{fig:par-rates}(Color online) Temperature dependences of ZF $\mu^+$ spin relaxation rates in the paramagnetic phase of $\alpha$-YbAl$_{0.75}$Fe$_{0.25}$B$_4$. (a)~Static Voigtian KT Gaussian rate~$\Delta_\mathrm{PM}$. (b)~Static Voitgtian KT exponential rate~$\lambda_\mathrm{PM}$ and dynamic rate~$\lambda_d$.}
\end{figure}

The temperature dependences of the ZF rates~$\Delta_\mathrm{PM}$, $\lambda_\mathrm{PM}$, and $\lambda_d$ in $\alpha$-YbAl$_{0.75}$Fe$_{0.25}$B$_4$ are shown in Fig.~\ref{fig:par-rates}. 
At 10~K and above all three quantities are essentially temperature-independent. The average value~$\Delta_\mathrm{PM}^{(\mathrm{av})} = 0.290(2)~\mu\mathrm{s}^{-1}$ is somewhat smaller than in $\alpha$-YbAl$_{0.986}$Fe$_{0.014}$B$_4$. This and the substantial value of $\lambda_\mathrm{PM}^{(\mathrm{av})}$ [$0.091(5)~\mu\mathrm{s}^{-1}$] can be attributed to the dilution of the $^{27}$Al nuclear spins by Fe substitution, which reduces the nuclear dipolar fields at $\mu^+$ sites and renders their distribution less Gaussian with more weight in the wings.

The dynamic rate~$\lambda_d$ is essentially zero over the entire temperature   range, in contrast to the nonzero rate in $\alpha$-YbAl$_{0.986}$Fe$_{0.014}$B$_4$.  This indicates that the spin fluctuation rate is significantly faster in $\alpha$-YbAl$_{0.75}$Fe$_{0.25}$B$_4$. In particular, above 10~K $\lambda_d$ does not exhibit the increase with decreasing temperature characteristic of critical slowing down. An increase is observed below $\sim$9~K, however (Sec.~\ref{sec:trans}, Fig.~\ref{fig:mag-rates}).

\subsubsection{Antiferromagnetic Phase} \label{sec:AFM}

The local field due to magnetic order in a crystal is expected to point in a well-defined crystalline direction, and thus may not be randomly oriented in a polycrystal if the latter is preferentially oriented~\cite{[{Local fields from nonrandomly-oriented dipoles with cubic site symmetry can yield a ZF KT relaxation function with late-time recovery~$G(t{\to}\infty) = 1/3$; see, e.g., }] CGRS77}. The $\alpha$-YbAl$_{0.75}$Fe$_{0.25}$B$_4$ sample is a mosaic of flat millimeter-sized single crystals glued to a silver plate. The crystalline $c$ axes are normal to the flat faces, and are therefore oriented preferentially along the initial $\mu^+$ spin direction. Preferential orientation changes ZF static KT relaxation functions in polycrystalline samples for $\mu^+$ sites with lower than cubic symmetry, principally by modifying the late-time constant $\mu^+$ spin polarization from the value~1/3 found for random orientation~\cite{Solt95,YaDdR11}. 

\begin{figure}[ht]
\includegraphics[clip=,width=3.25in]{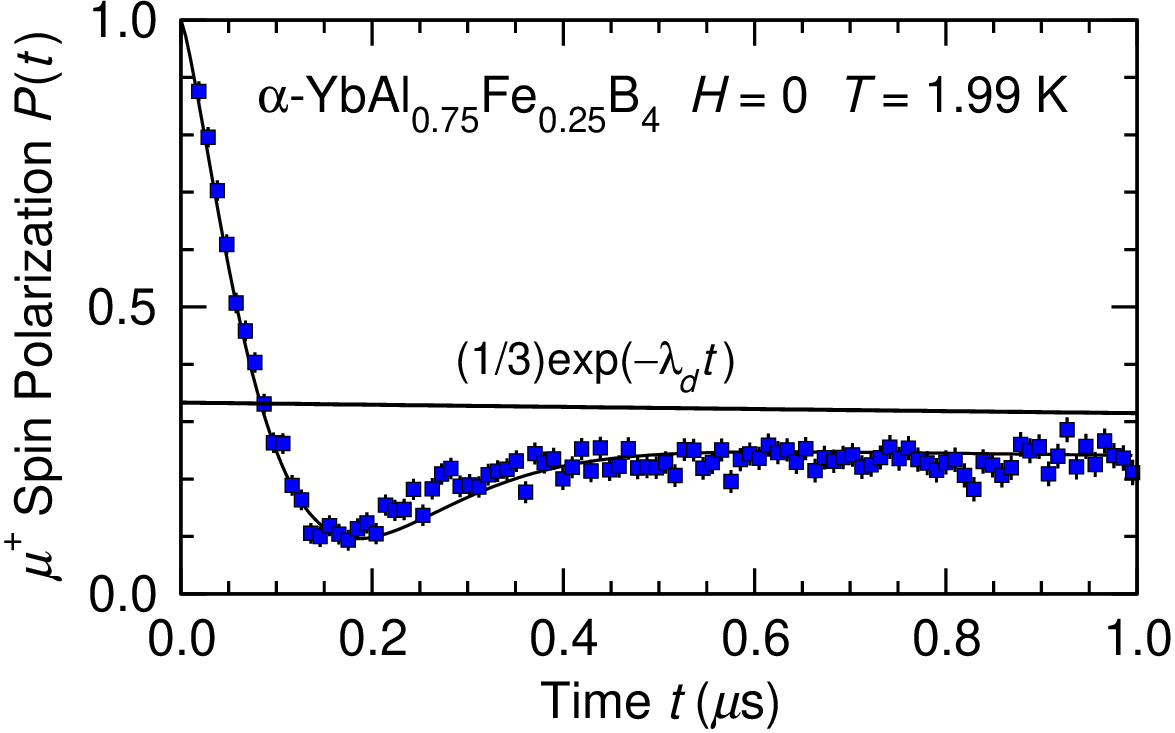}%
\caption{\label{fig:mag-offset}(Color online) $\alpha$-YbAl$_{0.75}$Fe$_{0.25}$B$_4$ $\mu^+$ spin polarization~$P(t)$ in ZF at $T = 1.99$~K\@. Preferential crystallite orientation reduces the amplitude of the slowly-relaxing polarization at late times from the value 1/3 expected for randomly-oriented local fields.}
\end{figure}

Figure~\ref{fig:mag-offset} gives the ZF $\mu^+$ spin polarization at 1.99~K\@. 
The data have been fit using an exponentially-damped ``offset-PE'' KT function
\begin{equation} \label{eq:mag-relax}
P_\mathrm{AFM}(t) = e^{-\lambda_d t}G'_\text{PE}(t) \,,
\end{equation}
where
\begin{equation} \label{eq:o-peKT}
G'_\text{PE}(t) = (1 - f_\mathrm{late})\left[3G_\text{PE}(t) - 1\right]/2 +  f_\mathrm{late} \,,
\end{equation}
with $G_\text{PE}(t)$ given by Eq.~(\ref{eq:edzfskt}). Equation~(\ref{eq:o-peKT}) simply replaces the constants 2/3 and 1/3 in Eq.~(\ref{eq:edzfskt}), appropriate to randomly-oriented fields, by $1 - f_\mathrm{late}$ and $f_\mathrm{late}$, respectively. It is a rough approximation for small $|f_\mathrm{late} - 1/3|$ to the exact result for preferential orientation assuming a uniaxial orientation distribution~\cite{Solt95,YaDdR11}.  The fit value of $f_\mathrm{late}$ is $0.22 < 1/3$, which indicates that the static $\mu^+$ internal fields are preferentially oriented perpendicular to the crystalline $c$ axes~\cite{Solt95}.

Parameters from damped offset-PE KT function fits to data in the 2--8~K temperature range are shown in Fig.~\ref{fig:mag-rates} and discussed in the next section. 

\subsubsection{Transition region} \label{sec:trans}

Magnetization measurements indicate multiple phase transitions in $\alpha$-YbAl$_{0.75}$Fe$_{0.25}$B$_4$ over the temperature range~6.9--9.4~K~\cite{SIKN14a}, but the data do not determine whether or not the various phases are macroscopically segregated. $\mu$SR is an ideal technique to probe inhomogeneous magnetism due to its sensitivity to static electronic magnetism, ordered or disordered.

As previously noted, fits of either the Voigtian or the PE function to the data over the entire temperature range give very poor fits in the transition region, suggesting an inhomogeneous distribution of transition temperatures. Magnetic resonance probes are sensitive to spatial distributions of local magnetism, ordered or disordered, if the correlation length~$\xi_M$ that describes this distribution is long enough so that each muon or nucleus is coupled to only one ``domain'' of the distribution. This usually means $\xi_M$ must be longer than a few lattice parameters.

\begin{figure}[ht]
\includegraphics[clip=,width=3.25in]{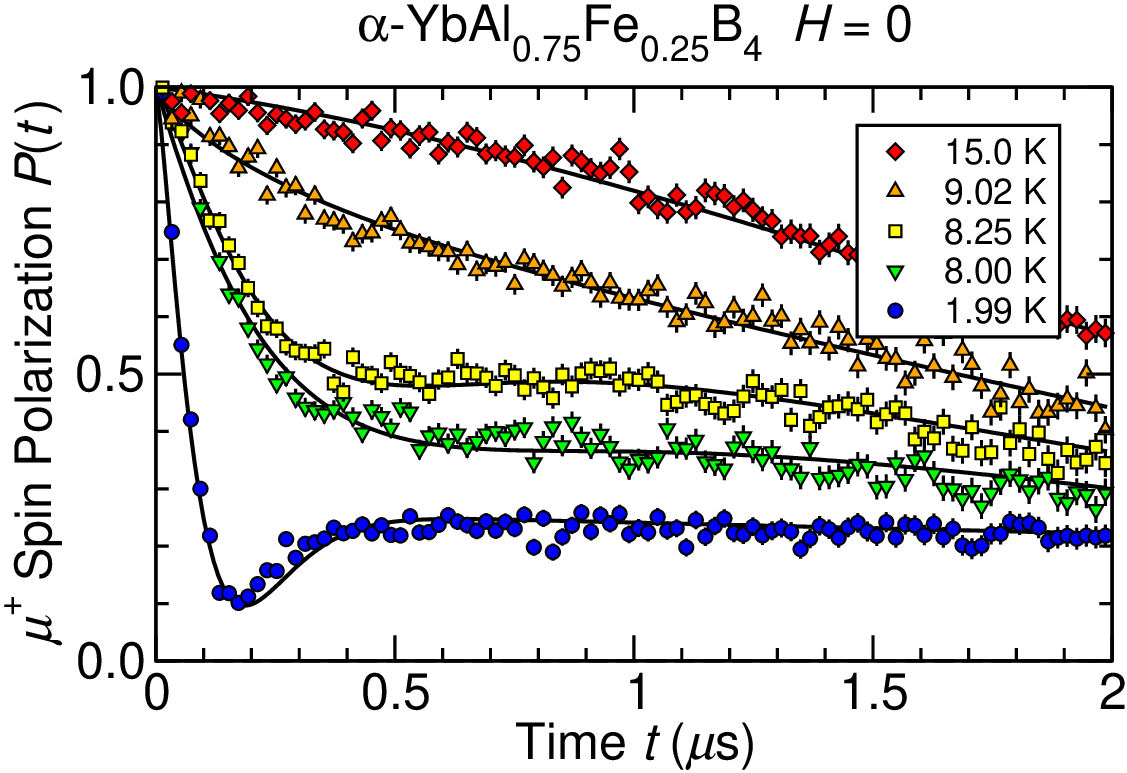}%
\caption{\label{fig:par-trans}(Color online) ZF $\mu^+$ spin polarization relaxation in $\alpha$-YbAl$_{0.75}$Fe$_{0.25}$B$_4$ at representative temperatures over the temperature range~2 K--15 K\@. The data exhibit tempera\-ture-dependent AFM and PM fractions in the AFM-PM transition region 7.5--10~K\@. Curves: fits to Eq.~(\ref{eq:trans}).}
\end{figure}

\begin{figure}[ht]
\includegraphics[clip=,width=3.25in]{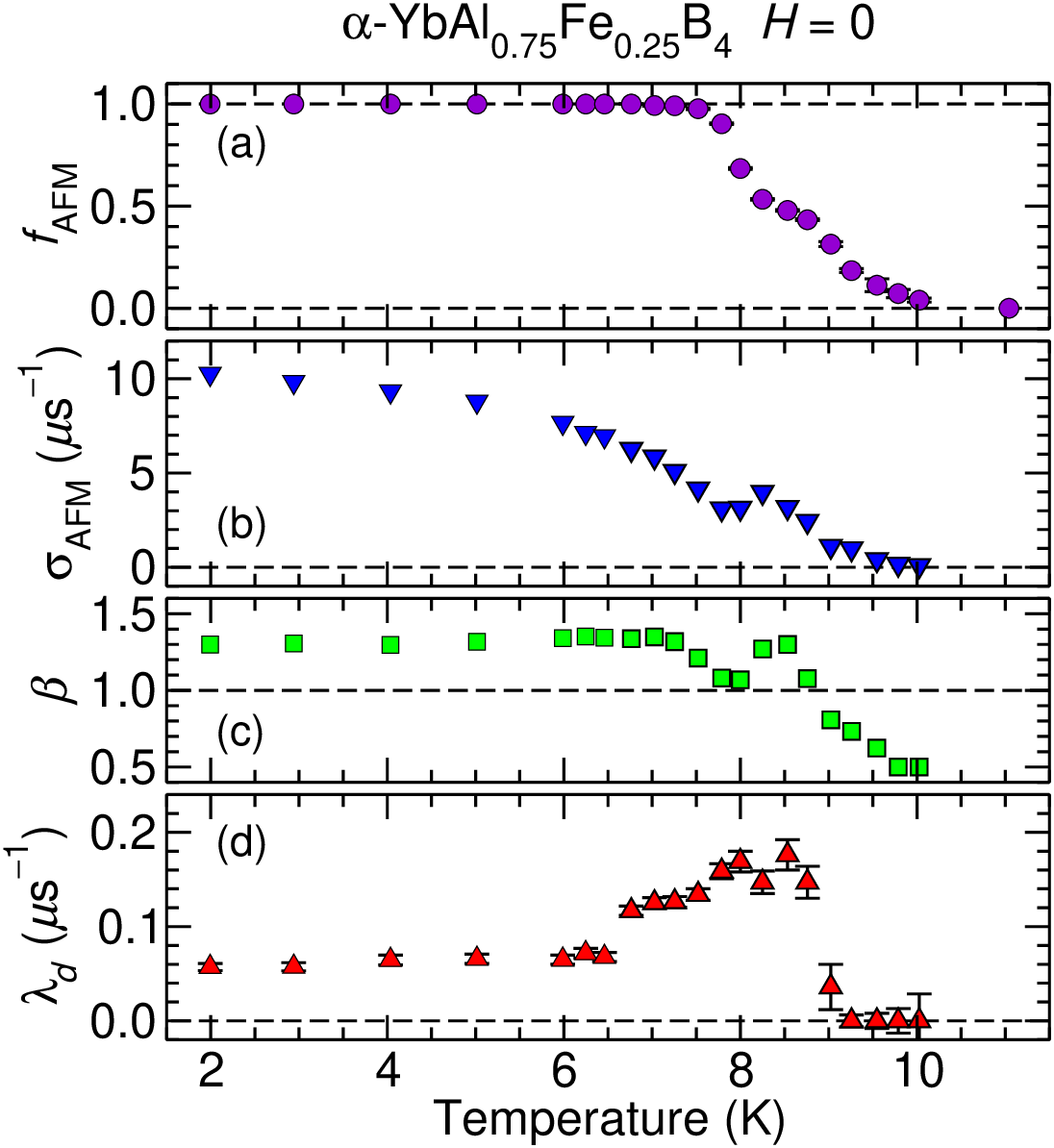}%
\caption{\label{fig:mag-rates}(Color online) Temperature dependences of AFM-phase $\mu^+$ spin relaxation parameters in the AFM and transition regions of $\alpha$-YbAl$_{0.75}$Fe$_{0.25}$B$_4$ from fits to Eqs.~(\ref{eq:mag-relax}) and (\ref{eq:trans}). (a)~Antiferromagnetic fraction~$f_\mathrm{AFM}$. (b)~Power-exponential relaxation rate~$\sigma_\mathrm{AFM}$. (c)~Exponent~$\beta$. (d)~Dynamic relaxation rate~$\lambda_d$.}
\end{figure}

The simplest assumption for such meso- or macroscopic inhomogeneity is a two-component (AFM and PM) form
\begin{equation} \label{eq:trans}
P(t) = f_\mathrm{AFM}P_\mathrm{AFM}(t) + (1 - f_\mathrm{AFM})P_\mathrm{PM}(t) \,,
\end{equation}
where $f_\mathrm{AFM}$ is the fraction of AFM phase. This scenario provides good fits over the entire temperature range, as shown in Fig.~\ref{fig:par-trans}.
The temperature dependences of $f_\mathrm{AFM}$ and the AFM-phase component parameters~$\sigma_\mathrm{AFM}$, $\beta$, and $\lambda_d$ from fits of Eqs.~(\ref{eq:mag-relax}) and (\ref{eq:trans}) to the data below $\sim$10~K are given in Fig.~\ref{fig:mag-rates}. 
In the fits $f_\mathrm{late}$ in Eq.~(\ref{eq:o-peKT}) has been fixed at its low-temperature value. The parameters of the PM-phase component~$P_\mathrm{PM}(t)$ in  Eq.~(\ref{eq:trans}) have been assumed temperature independent, and are fixed at their averages from data for $T \geqslant 10$~K (Fig.~\ref{fig:par-rates}).

It can be seen that $f_\mathrm{AFM}$ decreases monotonically over the transition region, suggesting a distribution of transition temperatures. There is, however, considerable structure in the temperature dependences of all the parameters, which we compare with the previously-reported transition temperatures~\cite{SIKN14a} $T_{N1} = 9.4(2)$~K, $T_{N2} = 8.0(2)$~K, and $T_{N3} = 6.9(1)$~K\@. 

(1)~From Fig.~\ref{fig:mag-rates}(a), with decreasing temperature $f_\mathrm{AFM}$ becomes nonzero below 10~K rather than $T_{N1}$.  There are inflection points in $f_\mathrm{AFM}(T)$ near 9~K and 8~K and saturation at $f_\mathrm{AFM} = 1$ below 7.5~K, i.e., no structure at 7~K\@. 

(2)~Recalling that the AFM-phase PE relaxation rate~$\sigma_\mathrm{AFM}$ [Fig.~\ref{fig:mag-rates}(b)] measures the strength of static fields (in frequency units), the decrease of $\sigma_\mathrm{AFM}(T)$ with increasing temperature from 2~K to 8~K is expected; it is the temperature dependence of the order parameter in this region. The minimum at $\sim$8~K and maximum at $\sim$8.3~K suggest structural transitions in the spin order. They might be associated with an increase in magnetic volume fraction, since $f_\mathrm{AFM}$ exhibits small additional increases with decreasing temperature at these temperatures [Fig.~\ref{fig:mag-rates}(a)]. Above 9~K $\sigma_\mathrm{AFM}$ becomes small but remains nonzero as long as $f_\mathrm{AFM} > 0$.

(3)~The exponent~$\beta$ [Fig.~\ref{fig:mag-rates}(c)] decreases from its low-temperature value above $\sim$7~K, goes through a minimum near 8~K and a maximum near 8.5~K, and decreases to less than 1 above $\sim$8.8~K\@. As noted above, a decrease of $\beta$ indicates broadening of the wings of the field distribution, i.e., increasing probability of field values far from the median. Not only is the volume fraction of the AFM phase decreasing with increasing temperature, but the disorder within this volume is increasing.

(4)~The dynamic rate~$\lambda_d$ [Fig.~\ref{fig:mag-rates}(d)] increases rapidly with decreasing temperature below $\sim$9~K to a poorly-defined maximum at 8--8.5~K, and then decreases to a constant value ${\sim}0.06~\mu\mathrm{s}^{-1}$ from $\sim$6~K down to 2~K\@. The maximum in the transition region suggests critical slowing down of spin fluctuations associated with the AFM transition. Of the parameters shown in Fig.~\ref{fig:par-rates} only $\lambda_d$ exhibits structure near $T_{N3}$ ($\sim$7~K)~\cite{SIKN14a}, below which it drops suddenly with decreasing temperature.

\section{CONCLUSIONS}

We have carried out $\mu$SR experiments on Fe-doped YbAlB$_4$ as a probe of quantum criticality and magnetic order in this alloy series. The principal results of this study are as follows.

For $x = 0.014$ there is no evidence of static magnetism, ordered or disordered. The dynamic muon spin relaxation rate~$\lambda_d$ exhibits a power-law temperature dependence~$\lambda_d \propto T^{-a}$, $a = 0.40(4)$, in the temperature range~100~mK--2~K\@. This divergence is similar to that found in materials with a putative FM QCP, and is in strong disagreement with predictions by theories of quantum critical behavior due to either AFM or valence fluctuations. With decreasing temperature $\lambda_d$ passes through a broad maximum at $\sim$50~mK, which might restore agreement with predicted valence critical behavior at lower temperatures, but the divergence above 100~mK would then remain unexplained. Further studies are necessary to clarify this situation.

For $x = 0.25$ the AFM state is inhomogeneous, with a broad distribution of local fields at $\mu^+$ sites and no indication of a well-defined average field. The inhomogeneity increases in the temperature region~7.5--10~K, where the $\mu$SR data indicate the coexistence of magnetically ordered and paramagnetic phases. This is evidence that the scale of the inhomogeneity is meso- or macroscopic, since otherwise each muon would sample both phases and the relaxation function would not exhibit the two-component behavior described in Sec.~\ref{sec:trans}. It is possible that Fe substitution is not random, so that clustering leads to a distribution of phase transition temperatures. There is evidence for a number of phase transitions from magnetization and $\mu$SR experiments, with rough but not perfect agreement between the transition temperatures. 

\begin{acknowledgments}
We are grateful to R. Abasalti, D. Arseneau, B. Hitti, S. Kreitzman, I. McKenzie, and G.~D. Morris of the TRIUMF Centre for Molecular and Materials Science for their help during these experiments, to A. Bianchi and A. Desilets-Benoit for useful discussions, and to Hu Cao for help with the experiments and data analysis. This work was supported in part by the U.S. National Science Foundation, grant nos.\ 0801407 (Riverside), 1105380 (Los Angeles), and 1104544 (Fresno), by the U.C. Riverside Academic Senate Committee on Research, by the National Natural Science Foundation of China (No.~11474060) and STCSM of China (No.~15XD1500200) (Shanghai), and by a Grant-in-Aid (No.~21684019) from the Japanese Society for the Promotion of Science (JSPS) and Grants-in-Aid for Scientific Research on Priority Areas (Nos.~17071003 and 19052003) from the Ministry of Education, Culture, Sports, Science and Technology (MEXT) (Kashiwa).
\end{acknowledgments}


%

\end{document}